\documentclass[aps,pre,twocolumn,superscriptaddress]{revtex4}
\usepackage[dvips]{graphics}
\usepackage{graphicx}
\usepackage{amsfonts}
\usepackage{amssymb}
\usepackage{amsmath}
\usepackage{subfigure}
\usepackage{color}

\newcommand{\RR}{\ensuremath{\mathbb R}} 
\newcommand{\CC}{\ensuremath{\mathbb C}}

\begin{document}

\title{Quantization of Spin Direction for Solitary Waves in a Uniform Magnetic Field}
\author{Q.E. Hoq}
\affiliation{Department of Mathematics, Western New England College, Springfield, MA,
01119, USA}

\begin{abstract}
It is known that there are nonlinear wave equations with localized solitary wave solutions. 
Some of these solitary waves are stable (with respect to a small perturbation of initial data) 
and have nonzero spin (nonzero intrinsic angular momentum in the center of momentum frame). 
In this paper we consider vector-valued solitary wave solutions to a nonlinear Klein-Gordon 
equation and investigate the behavior of these spinning solitary waves under the influence 
of an externally imposed uniform magnetic field. We find that the only  stationary spinning 
solitary wave solutions have spin parallel or anti-parallel to the magnetic field direction.
\medskip

\medskip

\end{abstract}

\thanks{E-mail address: qhoq@wnec.edu.}

\maketitle

%\email{qhoq@wnec.edu}

\section{Introduction}

Stable localized solitary wave solutions of nonlinear wave equations are known to exist 
(\cite{HW88}, \cite{Esteban}, \cite {HW95}, \cite{HW97}, \cite{HW98}, \cite{King}, \cite{HW02}). 
In particular, the existence of stable solitary wave solutions has been proven for certain nonlinear 
Klein-Gordon and Schr{\"o}dinger equations (\cite{Esteban}, \cite{HW95}, \cite{HW98}, \cite{HW02}) 
and it has also been established that these solutions can have nonzero spin (intrinsic angular momentum) 
(\cite{Esteban}, \cite{HW02}).

Here we consider the nonlinear Klein-Gordon equation (NLKG) 
\begin{equation}\label{NLKG1}
u_{tt}-\Delta u=\vec{g}(u)
\end{equation}
where $u:\RR^{3+1}\rightarrow\RR^{N}$ with N even (see \cite{HW98}) and the nonlinearity 
$\vec{g}:\RR^{N}\rightarrow \RR^{N}$ is defined by $\vec{g}(y)=g(|y|)\frac{y}{|y|}$,
for $y \neq 0$, $g:[0,\infty)\rightarrow \RR$ being a continuous function and $\vec{g}(0) = 0$.
%The function $h$ will be assumed to have the same properties as those listed 
%for the nonlinearity in \cite{HW95}, that is $h$ is locally-Lipschitz continuous with 
%$-\infty<-\sigma^2 \equiv\lim_{s \to 0} (h(s)/s)\leq 0$, 
%and $h(s)<0$ for small positive $s$ when $\sigma=0$. The primitive of $h$ will be assumed to have one 
%positive zero, with $h(s)>0$ for $s$ larger than this zero. 

In this paper we will examine the Noether conserved quantity 
\begin{eqnarray}
\vec{S}[u]\equiv\int_{\RR^3}u_t\cdot (\vec{X}\times\vec{\nabla}u)d^3\vec{X}.
\end{eqnarray}
called spin, which results from the rotational invariance of NLKG. (We note that 
for a function $u:\RR^{3+1}\longrightarrow\RR^N$, we define a counterclockwise 
rotation of $u$ about an axis through the origin in $\RR^3$ through an angle $\theta$ 
to be $u(R_\theta^{-1} \vec{X},t)$ where $R_{\theta}^{-1}$ is a $3\times 3$ 
counterclockwise rotation matrix.). This functional, $\vec{S}$, gives the angular 
momentum about the origin of a solution $u$. Our goal will be to find the spin of 
stationary solitary waves when exposed to an external uniform magnetic field, $\vec{B}$. 
The notation $S_x, S_y,$and $S_z$ will be used to denote the magnitude of the {\it x, y,}
and {\it z} components of $\vec{S}$ respectively.

Equation (\ref{NLKG1}) can be written compactly using relativistic index notation as 
\begin{equation}\label{NLKG2}
\partial^\alpha\partial_\alpha u=\vec{g}(u).
\end{equation}
where 
$\partial_\alpha=(\frac{\partial}{\partial{x}},\frac{\partial}{\partial{y}},\frac{\partial}{\partial{z}},\frac{\partial}{\partial{t}})$, 
and 
$\partial^\alpha=(-\frac{\partial}{\partial{x}}, -\frac{\partial}{\partial{y}},-\frac{\partial}{\partial{z}}, \frac{\partial}{\partial{t}})$. 
We model the imposition of an external uniform magnetic field of strength $B=|\vec{B}|$ 
parallel to the z-axis by making the minimal-coupling substitutions 
$\partial^{\alpha}\longmapsto \partial^{\alpha}-\sigma A^{\alpha}$ and $\partial_{\alpha}\longmapsto \partial_{\alpha}-\sigma A_{\alpha}$ into (\ref{NLKG2}) giving us
\begin{equation}\label{NLKGB}
(\partial^{\alpha}-\sigma{A^\alpha})(\partial_{\alpha}-\sigma{A_\alpha})u = \vec{g}(u)
\end{equation}
where $\sigma$ is a fixed $N\times N$ real skew-symmetric matrix with $\sigma^2=-I$ and
\begin{eqnarray}A=\frac{B}{2}\left(\begin{array}{c} -y \\ x \\ 0 \\ 0 \end{array}\right)
\end{eqnarray}
It will be assumed throughout that $B$ is small. This will allow us to simplify matters by 
ignoring terms in (\ref{NLKGB}) that involve $B^2$. We are going to look for stationary 
solitary waves solutions of (\ref{NLKGB}) that carry nonzero spin. A class of such solitary 
waves will be examined and it will be shown that they have spin either parallel or 
antiparallel to the uniform magnetic field. 

We look for standing-wave solutions of the form $u(\vec{X},t)=e^{t\Omega}v(\vec{X})$ 
where $\Omega$ is some $N\times N$ skew-symmetric matrix that commmutes with $\sigma$. 
We are interested in solutions that have $v(\vec{X})=\hat{\Psi}(\hat{X})w(r)$, where 
$w[0,\infty)\longrightarrow\RR$, with $r=(x^2+y^2+z^2)^{\frac{1}{2}}$, and 
$\hat{\Psi}:\RR^{3+1}\longrightarrow \RR^N$ is a unit-vector-valued eigenfunction of 
the spherical Laplacian. The Laplacian $\Delta$ can be decomposed into radial and 
angular components $\Delta = \Delta_R + \frac{1}{r^2}\Delta_S$. The spherical component 
$\Delta_S$, is a second-order derivative operator with only angular derivatives that acts 
on real-valued functions defined on the unit sphere. It also acts on real-valued functions 
defined on $\RR^3$, leaving the radial dependence unchanged. It is known that there exist 
unit-vector-valued eigenfunctions of $\Delta_S$ with any of the eigenvalues $\mu_l = -l(l+1)$, 
where $l = 0, 1, 2, 3, ...$ (\cite{HW98}, \cite{Folland}). Given a nonnegative integer 
{\it l}, such an eigenfunction $\hat{\Psi} : S^2 \longrightarrow S^{N-1}$ with eigenvalue 
$\mu_l$ exists provided that $N \geq 2l+1$. So $\Delta_S \hat{\Psi} = -l(l+1)\hat{\Psi}$, 
where we extend the action of $\Delta_S$ to vector $\RR^N$-valued functions by allowing 
the operator to act componentwise. Thus the coordinate functions of $\hat{\Psi}$ are 
eigenfunctions of $\Delta_S$. Substituting the ansatz 
\begin{equation}\label{ansatz}
u(\vec{X},t)=e^{t\Omega}\hat{\Psi}(\hat{X})w(r)
\end{equation}
into (1) produces the ordinary differential equation 
\begin{equation}\label{diffeq1}
w''(r)+\frac{2}{r}w'(r)-\frac{l(l+1)}{r^2} w(r)+f(w(r)) = 0
\end{equation}
where $l=0, 1, 2, 3, ...$ and $f(y) \equiv g(y) + \omega^2 y$ with appropriate conditions on 
{\it f} to ensure the existence of localized radial soutions (see \cite{HW98}). It is shown 
in \cite{HW98} that there then exist functions $w[0,\infty)\longrightarrow\RR$ that are 
exponentially decreasing far from the origin. 
%and that satisfy (6) with the initial conditions
%\begin{equation}
%\lim_{r \to 0^+}\frac{w(r)}{r^m}=d \qquad \text{and} \qquad \lim_{r \to 0^+}\bigl(\frac{w'(r)}{r^{m-1}}\bigl)=md
%\end{equation}
Consequently, for such functions $w$, $u(\vec{X},t)=e^{t\Omega}\hat{\Psi}(\hat{X})w(r)$ 
is a solution to NLKG. 

%It should be noted that $g(w(r))+\omega^2 w(r)$ satifies the same hypotheses as $g(w)$.

We will prove the following theorems:
\vspace{0.2in}

{\bf Theorem 1.} 
{\it Let B be a fixed real number, $\sigma$ an $N\times N$ real skew-symmetric matrix, 
and $\vec{g}:\RR^N\rightarrow\RR^N$ where $\vec{g}(v)=g(|v|^2)\frac{v}{|v|}$ with $\vec{g}(0) = 0$ for some 
continuous function $g:[0, \infty)\rightarrow\RR$. Let $u:\RR^{3+1}\rightarrow\RR^N$ 
be a solution of 
\begin{equation}
u_{tt}-\Delta u-B\sigma (x\partial_y-y\partial_x)=\vec{g}(u)
\end{equation}
of the form $u(\vec{X},t)=e^{t\Omega}\hat{\Psi}(\hat{X})w(r)$ where $\sigma$ is an $N\times N$ 
real skew-symmetric matrix with the property $[\Omega, \sigma]=0$, $\hat{\Psi}$ is a unit-vector-valued 
eigenfunction of the spherical Laplacian and {\it w} is a solution to (\ref{diffeq1}) as described above. 
Define the spin functional $\vec{S}$ by 
\begin{eqnarray}
\vec{S}[u]\equiv\int_{\RR^3}u_t\cdot (\vec{X}\times\vec{\nabla}u)d^3\vec{X}.
\end{eqnarray}
If $B\neq 0$ then $S_x[u]=S_y[u]=0$.}
\vspace{0.2in}

{\bf Theorem 2.}
{\it There are solutions with spin parallel (anti-parallel) to $\vec{B}$}.
\vspace{0.2in}

In section II we prove Theorems 1 and 2. Simply put, Theorem 1 states that solutions with 
the ansatz (\ref{ansatz}) to equation (\ref{NLKGB}) cannot have nonzero spin components 
that are not parallel to the magnetic field unless $\vec{B}=0$, while Theorem 2 imples 
that the only solutions of the ansatz to the equation have nonzero spin parallel to the 
$\vec{B}$ field. 

It is important to note that the discussion here is not a quantum mechanical one. Although 
many of the constructions have analogues in quantum mechanics, the interpretations are different.

\section{Proofs Of Theorems 1 And 2}
In what follows below, we will denote the {\it x, y}, and {\it z} components of 
$(-i)(\vec{X} \times \vec{\nabla})$ by $L_x$, $L_y$ and $L_z$ respectively. Thus
we have:
\begin{eqnarray}
L_x = -i\left( y\frac{\partial}{\partial z} - z\frac{\partial}{\partial y} \right) \\
L_y = -i\left( z\frac{\partial}{\partial x} - x\frac{\partial}{\partial z} \right) \\
L_z = -i\left( x\frac{\partial}{\partial y} - y\frac{\partial}{\partial x} \right)
\end{eqnarray}

Inserting (\ref{ansatz}) into (\ref{NLKGB}) gives
\begin{eqnarray}\label{expansion}
\Bigl[\Omega^2\hat{\Psi}(\hat{X})-B\sigma (x\partial_y-y\partial_x)\hat{\Psi}(\hat{X})\Bigl]w(r)- \notag \\ 
\Bigl(\Delta_Rw(r)-\frac{l(l+1)}{r^2}w(r)+g(w(r))\Bigl)\hat{\Psi}(\hat{X})=0
\end{eqnarray}
\vspace{0.2in}

Proof of {\bf Theorem 1}: Since both $\Omega$ and $\sigma$ are skew-symmetric, then
they are both normal. By hypothesis, $[\Omega, \sigma]=0$. Thus there exists an 
orthonormal basis of $\CC^N$ consisting of vectors which are eigenvectors for
both $\Omega$ and $\sigma$ (\cite{Nering}). Also since $\Delta_S$ and 
$L_z$ have a common set of orthonormal eigenfunctions, 
$\xi_{-l},...,\xi_{-1}, \xi_0, \xi_{1},...,\xi_{l}$ (\cite{Messiah})
that span the space of all eigenfunctions of $\Delta_S$ with eigenvalue $-l(l+1)$,
there exists $\alpha_{jm} \in \CC$, $-l \leq m \leq l$, $1 \leq j \leq N$, giving the expansion 
\begin{eqnarray}\label{psihat}
\hat{\Psi}(\hat{X})=\sum^N_{j=1}\sum^l_{m=-l}\alpha_{jm}\phi_j \xi_m(\hat{X})
\end{eqnarray}
where $\{\phi_j \}$ are an orthonormal basis of $\CC^N$ of eigenvectors of $\sigma$ and $\Omega$
with $\sigma \phi_j=\epsilon_j i \phi_j$ where $\epsilon_j=\pm 1$ and $\Omega\phi_j=i \nu_j \phi_j$ 
with real $\nu_j$. Substituting (\ref{psihat}) into (\ref{expansion}) we get 
\begin{eqnarray}
\sum^N_{j=1}\sum^l_{m=-l} \alpha_{jm} \phi_j \xi_m \Bigl\{\bigl(-\nu^2_j+B \epsilon_j m\bigl) w(r)-\eta (r) \Bigl\}=0
\end{eqnarray}
where $\eta \equiv \Delta_R w(r)-\frac{l(l+1)}{r^2}w(r)+g(w(r)).$ Therefore, for each $j$ and $m$ 
either 
\begin{eqnarray}\label{solve1}
\alpha_{jm}=0 
\end{eqnarray}
or
\begin{eqnarray}\label{solve2}
(-\nu^2_j+\epsilon_j Bm)w(r)-\eta(r)=0
\end{eqnarray}

Next we compute the quantity $S_x[u]+iS_y[u]$. It should be noted that since $u_t$ is real, 
it is equal to its complex conjugate. This will enable us to simplify things later. From
\cite{Messiah} we recall that $(L_x + iL_y)\xi_m = \sqrt{l(l+1)-m(m+1)} \xi_{m+1}$. Using this

\begin{eqnarray}
S_x[u]+iS_y[u] = \sum^N_{j=1}\sum^{l-1}_{m=-l}\Bigl( \int_0^{\infty} w^2(r)r^2 dr\Bigl) \times \notag \\
\overline{\alpha}_{j(p+1)} \alpha_{jp} \nu_j \sqrt{l(l+1)-p(p+1)}
\end{eqnarray}

If $S_x[u]+iS_y[u]$ is to be nonzero, then for some {\it m}, {\it j} and {\it p}, $\alpha_{j(p+1)}\neq 0$ 
and $\alpha_{jp}\neq 0$. From (\ref{solve1}) and (\ref{solve2}) then

\begin{eqnarray}
\Bigl( -\nu^2_j+\epsilon_j B(p+1) \Bigl) w(r)-\eta(r)=0
\end{eqnarray}
and
\begin{eqnarray}
\Bigl( -\nu^2_j+\epsilon_j Bp \Bigl) w(r)-\eta(r)=0
\end{eqnarray}

Subtract to get 

\begin{eqnarray}
Bw(r)=0 \\
\Rightarrow B=0 ~~~~\text{or} ~~~~w(r)\equiv 0
\end{eqnarray}

So if $B \neq 0$ then $S_x+iS_y=0$
\vspace{0.2in}

Proof of {\bf Theorem 2}: Without loss of generality, we assume $B>0$ and take 

\begin{eqnarray}
\sigma=
\left( \begin{array}{cccc} 
        0          &   -1          &   0                 & 0                     \\
        1          &    0          &   0                 & 0                     \\
        0          &    0          &   0                 &-1                     \\
        0          &    0          &   1                 &0                      \\
\end{array} \right)
\end{eqnarray}

It is straightforward to show that 

\begin{eqnarray}
\hat{\Psi} (\hat{X})\equiv \frac{1}{r}\left(\begin{array}{c} x \\ y \\ z \\ 0 \end{array}\right)
\end{eqnarray}
is an eigenfunction of $\Delta_s$ with eigenvalue $-l(l+1)=-2$. In (\ref{expansion})
we will choose $\Omega$ so that 

\begin{equation}
\Omega^2 \hat{\Psi}(\hat{X}) -B\sigma (x\partial_y-y\partial_x)\hat{\Psi}(\hat{X})
\end{equation}
is a multiple of $\hat{\Psi}$.
Let

\begin{eqnarray}
\Omega = 
\left( \begin{array}{cccc} 
        0          &   -\omega_1   &   0                 & 0                     \\
        \omega_1   &    0          &   0                 & 0                     \\
        0          &    0          &   0                 &-\omega_2     \\
        0          &    0          &\omega_2  &0                      \\
\end{array} \right)
\end{eqnarray}

Substituting into 
\begin{equation}
\Omega^2 \hat{\Psi}(\hat{X}) -B\sigma (x\partial_y-y\partial_x)\hat{\Psi}(\hat{X})
\end{equation}
produces 

\begin{eqnarray}
\left(\begin{array}{c} (-\omega_1^2+B)x \\ (-\omega_1^2+B)y \\ -\omega_2^2z \\ 0 \end{array}\right)
\end{eqnarray}
This will be a multiple of $\hat{\Psi}$ whenever $\omega_2=\pm \sqrt{\omega_1-B}$. So

\begin{eqnarray}
\Omega=
\left( \begin{array}{cccc} 
        0          &   -\omega     &   0                 & 0                     \\
        \omega     &    0          &   0                 & 0                     \\
        0          &    0          &   0                 &-\sqrt{\omega^2-B}     \\
        0          &    0          &\sqrt{\omega^2-B}    &0                      \\
\end{array} \right)
\end{eqnarray}

We put this form of $\Omega$ into the wave equation to get an ordinary differential equation for $w$: 

\begin{eqnarray}\label{diffeq2}
w''+\frac{2}{r}w'-\frac{2}{r^2}w+g(w)+(\omega^2-B)w=0
\end{eqnarray}

This is known to have exponentially localized solutions provided 
$\lim_{s \to 0}\frac{g(s)+(\omega^2-B)s}{s}\leq 0$ 
(\cite{HW98}). So we require $-\rho^2 +\omega^2 -B\leq 0$ (where $\lim_{s \to 0}\frac{g(s)}{s}=\rho^2$) 
which is the case for all $\omega$ with $0\leq \omega^2 \leq \rho^2+B$. Then, according 
to the results of \cite{HW98}, there are $C^2$ exponentially localized solutions 
to (\ref{diffeq2}). The corresponding functions $u(\vec{X},t)=e^{t\Omega}\hat{\Psi}(\hat{X})w(r)$ 
are solutions to (\ref{NLKGB}). 

Using the form of $\Omega$ we now show that a solution of the form 
$u(\overrightarrow{X},t)=e^{t\Omega}\hat{\Psi}(\hat{X})w(r)$ has nonzero spin parallel to 
the direction of the magnetic field $B$. The component of the spin parallel to $\vec{B}$ is
\begin{eqnarray}
S_z[u]&=&\int_{\RR^3}w^2 \Omega \hat{\Psi}(\hat{X})\cdot (x\partial_y-y\partial_x)\hat{\Psi}(\hat{X}) d^3\vec{X} \\
&=&\frac{\omega}{r^2}(y^2+x^2) \\
&> 0&
\end{eqnarray}
if $\omega>0$ (and both {\it x} and {\it y} are nonzero).
So,

\begin{eqnarray}
S_z[u]=\int_{\RR^3} \frac{\omega^2}{r^2}\omega (x^2+y^2) d^3\vec{X}
\end{eqnarray}
is nonzero.


\begin{thebibliography}{99}
\bibitem{HW88}   
   M. Deumens, H. Warchall. Explicit Construction Of All Spherically 
   Symmetric Solitary Waves For A Nonlinear Wave Equation In Multiple Spatial 
   Dimensions, {\it Nonlinear Anal.}, \textbf{12} 419-447 (1988).

\bibitem{Esteban}
    M. J. Esteban, P.-L. Lions, Stationary solutions of nonlinear Schrödinger 
    equations with an external magnetic field, {\it Partial differential equations 
    and the calculus of variations}, Vol. I, 401-449, Progr. Nonlinear Differential 
    Equations Appl., 1, (Birkhäuser Boston, Boston, MA, 1989).

\bibitem{HW95}
   J.A. Iaia, H. Warchall. Nonradial Solutions of a Semilinear 
   Elliptic Equation in Two Dimensions. {\it Journal Of Differential Equations} 
   $\bf{119}$ 533-558 (1995)

\bibitem{HW97}
   J. A. Iaia, H. A. Warchall, F. B. Weissler. Localized Solutions 
   of Sublinear Elliptical Equations: Loitering at the Hilltop. {\it Rocky 
   Mountain Journal of Mathematics} \textbf{27} 131-1157 (1997).

\bibitem{HW98}
   J.A. Iaia, H. Warchall. Encapsulated-Vortex Solutions To Equivariant Wave                     
   Equations:Existence. {\it SIAM J. Math. Anal.} \textbf{30} 118-139 (1998).

\bibitem{King}
   G. King. {\it Explicit Multidimensional Solitary Waves}, Master's Thesis, 
   University Of North Texas, Denton, TX (1990).

\bibitem{HW02}
   R. L. Pego, H. Warchall. Spectrally Stable Encapsulated Vortices For Nonlinear              
   Schr$\ddot{o}$dinger Equations. {\it J. Nonlinear Sci.} \textbf{12} 347-394 (2002)

\bibitem{Folland}
   G.B. Folland. {\it Introduction To Partial Differential Equations}.
   (Princeton University Press, 1995). 
    
\bibitem{Nering}
   E. D. Nering. {\it Linear Algebra and Matrix Theory}.
   (John Wiley and Sons, 1963)
    
\bibitem{Messiah}
   A. Messiah. {\it Quantum Mechanics}
   (John Wiley and Sons, New York, 1963).

%\bibitem{Strauss}
%  W. A. Strauss. {\it Nonlinear Wave Equations}. (Conference Board of the 
%  Mathematical   Sciences, Regional Coference Series in Mathematics, American Mathematical 
%  Society, Number 73).


\end{thebibliography}
\end{document}